\documentclass[11pt]{article}
\usepackage{graphicx}

\newcommand{\BABARPubYear}    {06}

\newcommand{\BABARConfNumber} {012}
\newcommand{\SLACPubNumber} {11975}

\input pubboard/babarsym
\RequirePackage{xspace}

\newcommand{\calP}{\ensuremath{{\cal P}}}

\newcommand{\pvec}{{\bf p}}

\providecommand{\skz}{\mbox{$S$}}
\providecommand{\ckz}{\mbox{$C$}}
\def\deltaS{\ensuremath{{\rm \Delta}S}\xspace}

\newcommand{\DE}{\ensuremath{\Delta E}}
\newcommand{\mb}{\ensuremath{m_{\rm ES}}}
\newcommand{\mres}{\ensuremath{m_{\rm res}}}
\newcommand{\xf}{\ensuremath{{\cal F}}}
\newcommand{\hel}{\ensuremath{{\cal H}}}

\providecommand{\dt}{\deltat}
\newcommand{\ttag}{\ensuremath{t_{\rm tag}}}
\newcommand{\bflav}{\ensuremath{B_{\rm flav}}}

\newcommand\etal{{\it et al.}}
\newcommand{\half}{\ensuremath{{1\over2}}}

\newcommand{\msp}{\ensuremath{\phantom{-}}}

\newcommand{\bfig}{\begin{figure}[htbpc!]}
\newcommand{\efig}{\end{figure}}
\newcommand\bef{\begin{figure}}
\newcommand\edf{\end{figure}}
\newcommand\dbline{\noalign{\vskip 0.10truecm\hrule}\noalign{\vskip 2pt}\noalign{\hrule\vskip 0.10truecm}}

\newcommand\sgline{\noalign{\vskip 0.10truecm\hrule\vskip 0.10truecm}}
\newcommand\beq{\begin{equation}}
\newcommand\eeq{\end{equation}}
\newcommand\bear{\begin{array}}
\newcommand\enar{\end{array}}
\newcommand\beqa{\begin{eqnarray}}
\newcommand\eeqa{\end{eqnarray}}
\newcommand\ben{\begin{enumerate}}
\newcommand\een{\end{enumerate}}

\newcommand{\UfourS}{\ensuremath{\Upsilon(4S)}}

\newcommand{\omtoppp}{\ensuremath{{\omega\ra\pip\pim\piz}}}

\newcommand{\fomegaKz}{\ensuremath{\omega K^0}}
\newcommand{\fomegaKs}{\ensuremath{\omega\KS}}
\newcommand{\omegaKz}{\ensuremath{\Bz\ra\fomegaKz}}
\newcommand{\omegaKs}{\ensuremath{\Bz\ra\fomegaKs}}

\newcommand{\SomegaKz}{\ensuremath{0.xx^{+0.xx}_{-0.xx}\pm zz}}
\newcommand{\ComegaKz}{\ensuremath{-0.xx^{+0.xx}_{-0.xx}\pm zz}}

\renewcommand{\mres}{\ensuremath{m_{3\pi}}}
\providecommand{\sigdt}{\ensuremath{\sigma_{\deltat}}}
\renewcommand{\SomegaKz}{\ensuremath{0.62^{+0.25}_{-0.30}\pm 0.02}}
\renewcommand{\ComegaKz}{\ensuremath{-0.43^{+0.25}_{-0.23}\pm 0.03}}

\long\def\inst#1{\par\nobreak\kern 4pt\nobreak
    {\it #1}\par\vskip 10pt plus 3pt minus 3pt}

\begin{document}
{\pagestyle{empty}

\begin{flushright}
\babar-CONF-\BABARPubYear/\BABARConfNumber \\
SLAC-PUB-\SLACPubNumber \\
\end{flushright}

\par\vskip 5cm

\begin{center}
\Large \bf Measurements of  $CP$-Violating Asymmetries
in $B$ Decays to $\omega \KS$
\end{center}
\bigskip

\begin{center}
\large The \babar\ Collaboration\\
\mbox{ }\\
\today
\end{center}
\bigskip \bigskip

\begin{center}
\large \bf Abstract
\end{center}
We present preliminary measurements of \CP-violating asymmetries
for the decay \omegaKs.  The data 
sample corresponds to 347 million \BB\ pairs produced by \epem\ annihilation 
at the \UfourS\ resonance.  
For the decay \omegaKs, we measure the time-dependent \CP-violation parameters
$\skz = \SomegaKz$, and $\ckz = \ComegaKz$, where the first uncertainty is 
statistical and the second systematic. 

\vfill
\begin{center}

Submitted to the 33$^{\rm rd}$ International Conference on High-Energy Physics, ICHEP 06,\\
26 July---2 August 2006, Moscow, Russia.

\end{center}

\vspace{1.0cm}
\begin{center}
{\em Stanford Linear Accelerator Center, Stanford University, 
Stanford, CA 94309} \\ \vspace{0.1cm}\hrule\vspace{0.1cm}
Work supported in part by Department of Energy contract DE-AC03-76SF00515.
\end{center}

\newpage
} 


\begin{center}
\small

The \babar\ Collaboration,
\bigskip

%
{B.~Aubert,}
{R.~Barate,}
{M.~Bona,}
{D.~Boutigny,}
{F.~Couderc,}
{Y.~Karyotakis,}
{J.~P.~Lees,}
{V.~Poireau,}
{V.~Tisserand,}
{A.~Zghiche}
\inst{Laboratoire de Physique des Particules, IN2P3/CNRS et Universit\'e de Savoie,
 F-74941 Annecy-Le-Vieux, France }
% ...

\end{center}\newpage

\section{INTRODUCTION}
\label{sec:Introduction}

Measurements of time-dependent \CP\ asymmetries in \Bz\ meson decays through
a Cabibbo-Kobaya\-shi-Maskawa (CKM) favored $b \to c \bar{c} s$ amplitude
\cite{s2b,belles2b} have firmly established that \CP\ symmetry is not 
conserved in the neutral $B$ meson system. 
The effect, arising from the interference between mixing and decay
proportional to the \CP-violating phase $\beta = \arg{(-V_{cd} V^*_{cb}/
V_{td} V^*_{tb})}$ of the CKM mixing matrix \cite{SM},
manifests itself as an asymmetry in the time evolution of the $\Bz\Bzb$ pair.

In the Standard Model, decays of $B^0$ mesons to charmless hadronic final states 
such as $\omega\Kz$ proceed mostly via a single loop (penguin) 
amplitude with the same weak phase as the $b \to c \bar{c} s$
transition \cite{Penguin}, but  
CKM-suppressed amplitudes and multiple particles in the loop introduce 
additional weak phases whose contribution may not be negligible; 
see Refs. \cite{Gross,BN} for early quantitative work in
addressing the size of these effects.  We define \deltaS\ as the difference 
between the magnitude of the time-dependent \CP-violating parameter $S$ 
(given in detail below) 
measured in these decays and $S=\stwob$ measured in decays to 
charmonium and a neutral kaon. 
For the decay \omegaKz, these additional contributions are expected to
give $\deltaS\sim$ 0.1 \cite{beneke,CCS}, although this increase may be 
nullified when final-state interactions are included \cite{CCS}.
A value of \deltaS\ inconsistent with this expectation could be an indication 
of new physics \cite{lonsoni}.

We present an improved preliminary measurement of the time-dependent
\CP-violating asymmetry in the decay \omegaKs, previously reported by the
\babar\ and Belle Collaborations~\cite{BABAR, BELLE}. Charge-conjugate decay 
modes are implied throughout.

\section{THE \babar\ DETECTOR AND DATASET}
\label{sec:babar}

The data were collected with the 
\babar\ detector~\cite{BABARNIM} at the PEP-II asymmetric-energy
 $e^+e^-$ collider.
An integrated luminosity of 316~fb$^{-1}$, corresponding to
347 million \BB\ pairs, was recorded at the $\Upsilon (4S)$
resonance (center-of-mass energy $\sqrt{s}=10.58\ \gev$).
Charged particles are detected and their momenta measured by the
combination of a silicon vertex tracker (SVT), consisting of five layers
of double-sided detectors, and a 40-layer central drift chamber,
both operating in a 1.5 T axial magnetic field.  Charged-particle 
identification 
is provided by the energy loss in the tracking devices and by the
measured Cherenkov angle from an internally reflecting ring-imaging
Cherenkov detector covering the central region.
Photons and electrons are detected by a CsI(Tl) electromagnetic calorimeter.
The instrumented flux return  of the magnet allows discrimination of
muons from pions.

\section{ANALYSIS METHOD}
\label{sec:Analysis}

From a $\Bz\Bzb$ pair produced in an \UfourS\ decay, we reconstruct one
of the $B$ mesons in the final state $f = \fomegaKs$, a \CP\ eigenstate
with eigenvalue $-1$.  For the time evolution measurement, we also
identify (tag) the flavor (\Bz\ or \Bzb) and reconstruct the
decay vertex of the other $B$.
The asymmetric beam configuration in the laboratory frame
provides a boost of $\beta\gamma = 0.56$ to the center-of-mass in the lab
frame, which
allows the determination of the proper decay time difference $\dt \equiv
t_f-\ttag$ from the vertex separation of the two $B$ meson candidates.
Ignoring the \dt\ resolution (about 0.5 ps), the distribution of \dt\ is
\begin{equation}
   \label{eq:FCPdef}
  F(\dt) = 
        \frac{e^{-\left|\deltat\right|/\tau}}{4\tau} [1 \mp\Delta w \pm
   (1-2w)\left( S\sin(\deltamd\deltat) -
C\cos(\deltamd\deltat)\right)].
\end{equation}
The upper (lower) sign denotes a decay accompanied by a \Bz (\Bzb) tag,
$\tau$ is the mean $\Bz$ lifetime, $\deltamd$ is the mixing
frequency, and the mistag parameters $w$ and
$\Delta w$ are the average and difference, respectively, of the probabilities
that a true $\Bz$\,($\Bzb$) meson is mistagged as a $\Bzb$\,($\Bz$).
The parameter $C$ measures direct \CP\ violation.  

The flavor-tagging algorithm \cite{s2b} has seven mutually exclusive tagging categories
of differing purities, including one for untagged events that we
retain for yield determinations.  The measured analyzing power, defined as 
efficiency times $(1-2w)^2$ summed over all categories, is $( 30.4\pm 0.3)\%$,
as determined from a large sample 
of $B$ decays to fully reconstructed flavor eigenstates (\bflav).

We reconstruct a $B$ meson candidate by combining a \KS\ with an 
\omtoppp\ candidate.  We select $\KS\to\pi^+\pi^-$ decays by requiring 
the $\pi^+\pi^-$ 
invariant mass to be within 12 \mev\ ($\sim$4$\sigma$) of the 
nominal \Kz\ mass and by
requiring a flight length greater than three times its error.  
We require the
$\pip\pim\piz$ invariant mass (\mres) to be between 735 and 825 \mev.  Distributions 
from the data and from Monte Carlo (MC) simulations \cite{geant}\ guide the 
choice of these selection criteria.  We retain regions adequate to
characterize the background as well as the signal for those quantities taken 
subsequently as observables for fitting.  We also use the angle 
$\theta_H$, defined in the $\omega$ rest frame as the angle of the
direction of the boost from the $B$ rest frame with respect to the normal to 
the $\omega$ decay plane.  The quantity $\hel\equiv|\cos{\theta_H}|$ is
approximately flat for background decays and distributed as 
$\cos^2{\theta_H}$ for signal decays.

A $B$ meson candidate is characterized kinematically by the energy-substituted mass
$\mes \equiv  \sqrt{(\half s + \pvec_0\cdot \pvec_B)^2/E_0^2 - \pvec_B^2}$ and the 
energy difference
$\DE \equiv E_B^*-\sqrt{s}/2$, where 
$(E_0,\pvec_0)$ and $(E_B,\pvec_B)$ are four-momenta of
the \UfourS\ and the $B$ candidate, respectively, and the asterisk 
denotes the center-of-mass rest frame.  We require
$|\DE|\le0.2$ GeV, $5.25\le\mes\le5.29\ \gev$, $|\dt|<20$ ps and $\sigdt<2.5$ ps.

To help reject the dominant background from continuum $\epem\ra\qqbar$ events 
($q=u,d,s,c$), we use
the angle $\theta_T$ between the thrust axis of the $B$ candidate and
that of
the rest of the tracks and neutral clusters in the event, calculated in
the  \UfourS\ rest frame.  The distribution of $\cos{\theta_T}$ is
sharply peaked near $\pm1$ for jet-like $q\bar q$
pairs and is nearly uniform for the isotropic $B$ decays; we require
$|\cos{\theta_T}|<0.9$ .

From MC simulations of \BzBzb\ and \BpBm\ events, we find evidence 
for a small (0.3\% of the total sample) \BB\ background contribution.
We have therefore added a \BB\ component to the fit described below.

We use an unbinned, multivariate maximum-likelihood fit to extract
signal yields and \CP-violation parameters.  We use the discriminating 
variables \mes, \DE, \mres, \hel, and a Fisher 
discriminant \xf\ \cite{PRD}. The Fisher discriminant combines five 
variables: the polar angles with respect to the beam axis in the
\UfourS\ frame of the $B$ candidate momentum and of the $B$ thrust axis; 
the tagging category; and the zeroth and second 
angular moments of the energy flow, excluding the $B$ candidate, about the 
$B$ thrust axis \cite{PRD}.  We use \dt to extract the \CP-violation parameters,
$S$ and $C$.

We define the probability density function
(PDF) for each event $i$, hypothesis $j$ (signal, \BB\ background 
and \qqbar\ background), and tagging category $c$:
\begin{equation}
\calP_{j,c}^i  \equiv \calP_j (\mes^i) \calP_j (\DE^i) \calP_j(\xf^i, c)
\calP_j (\mres^i)
\calP_j(\hel^i) \calP_j (\dt^i, \sigdt^i, c)\,, 
\end{equation}
where $\sigdt^i$ is the error on \dt\ for event $i$.  
We write the extended likelihood function as
\begin{equation}
{\cal L} = \prod_{c} \exp{(-\sum_j Y_{j} f_{j,c})}
\prod_i^{N_c}\left[\sum_j Y_j f_{j,c} {\cal P}^i_{j,c}\right]\,,
\end{equation}
where $Y_j$ is the fitted yield of events of species $j$, $f_{j,c}$ is the
fraction of events of species $j$ for each category $c$,
and $N_c$ is the number of events of category $c$ in the sample. 
We fix $f_{{\rm sig},c}$ and $f_{\BB,c}$ 
to $f_{\bflav,c}$, the values measured with the large
\bflav\ sample  \cite{s2b}.  

The PDF $\calP_{\rm sig}(\dt,\, \sigdt, c)$ is given by
$F(\dt)$ (Eq.\ \ref{eq:FCPdef}) with tag category ($c$) dependent mistag
parameters convolved with the
signal resolution function (a sum of three Gaussians) determined from the
\bflav\ sample.
The other PDF forms are: the sum of two Gaussians for all signal shapes
except \hel, and for the peaking component of the \mres\ background;
the sum of three Gaussians for 
$\calP_{\qqbar}(\dt, c)$ and $\calP_{\BB}(\dt, c)$; an asymmetric Gaussian 
with different
widths below and above the peak for $\calP_j(\xf)$ (a small ``tail"
Gaussian is added for $\calP_{\qqbar}(\xf)$); Chebyshev functions of
second to fourth order for the \hel\ distribution for signal and the 
slowly-varying shapes of the \DE, 
\mres, and \hel\ distributions for backgrounds; and, for 
$\calP_{\qqbar}(\mes)$, a
phase-space-motivated empirical function \cite{argus}, with a small Gaussian 
added for $\calP_{\BB}(\mes)$.

We determine the PDF parameters from simulation for the signal and \BB\
background components.  We study large control samples of $B\to D\pi$ decays 
of similar topology to verify the simulated
resolutions in \DE\ and \mes, adjusting the PDFs to account for any
differences found.  For the \qqbar\ background we use
(\mes,\,\DE) sideband data to obtain initial PDF-parameter values, but ultimately 
leave many of them free to vary in the final fit.

\section{RESULTS}
\label{sec:Physics}

The free parameters in the fit are 
the following: the signal, \BB\ background, and \qqbar\ background yields;
the three shape parameters of 
${\cal P}_{\qqbar}(\xf)$; the slopes of ${\cal P}_{\qqbar}(\DE)$ and
${\cal P}_{\qqbar}(\mres)$; the fraction of the peaking component of
${\cal P}_{\qqbar}(\mres)$; the \mes\ background shape parameter
$\xi$ \cite{argus}; $S$; $C$;
 the fraction of background events in each
tagging category; and the six primary parameters describing the \dt\
background shape.  The parameters $\tau$ and 
$\deltamd$ are fixed to world-average values \cite{PDG2006}.
Table \ref{tab:results} shows the results of the fit.
The errors have been scaled by $\sim$1.10 to account for a slight
underestimate of the fit errors predicted by our simulations when the
number of signal events is small.

\begin{table}[ht]
\caption{
Total sample size, detection efficiency, signal yield, \BB\ background 
yield and $CP$-asymmetry parameters $S$ and $C$ from the fit.}
\label{tab:results}
\begin{center}
\begin{tabular}{lccc}
\dbline
Quantity	    & \fomegaKs\ \\ 
\sgline
Total fit sample  &     12636     \\
Eff. (\%) & 23.0   \\
Fit signal yield  & $142^{+17}_{-16}$  \\
\BB\ yield        & $38^{+25}_{-22}$  \\
$S$                 &\msp$0.62^{+0.25}_{-0.30}$ \\
$C$              &  $-0.43^{+0.25}_{-0.23}$ \\
\dbline
\end{tabular}
\end{center}
\vspace*{-0.5cm}
\end{table}

\begin{figure}[!htb]
\begin{center}
 \includegraphics[angle=0,scale=0.7]{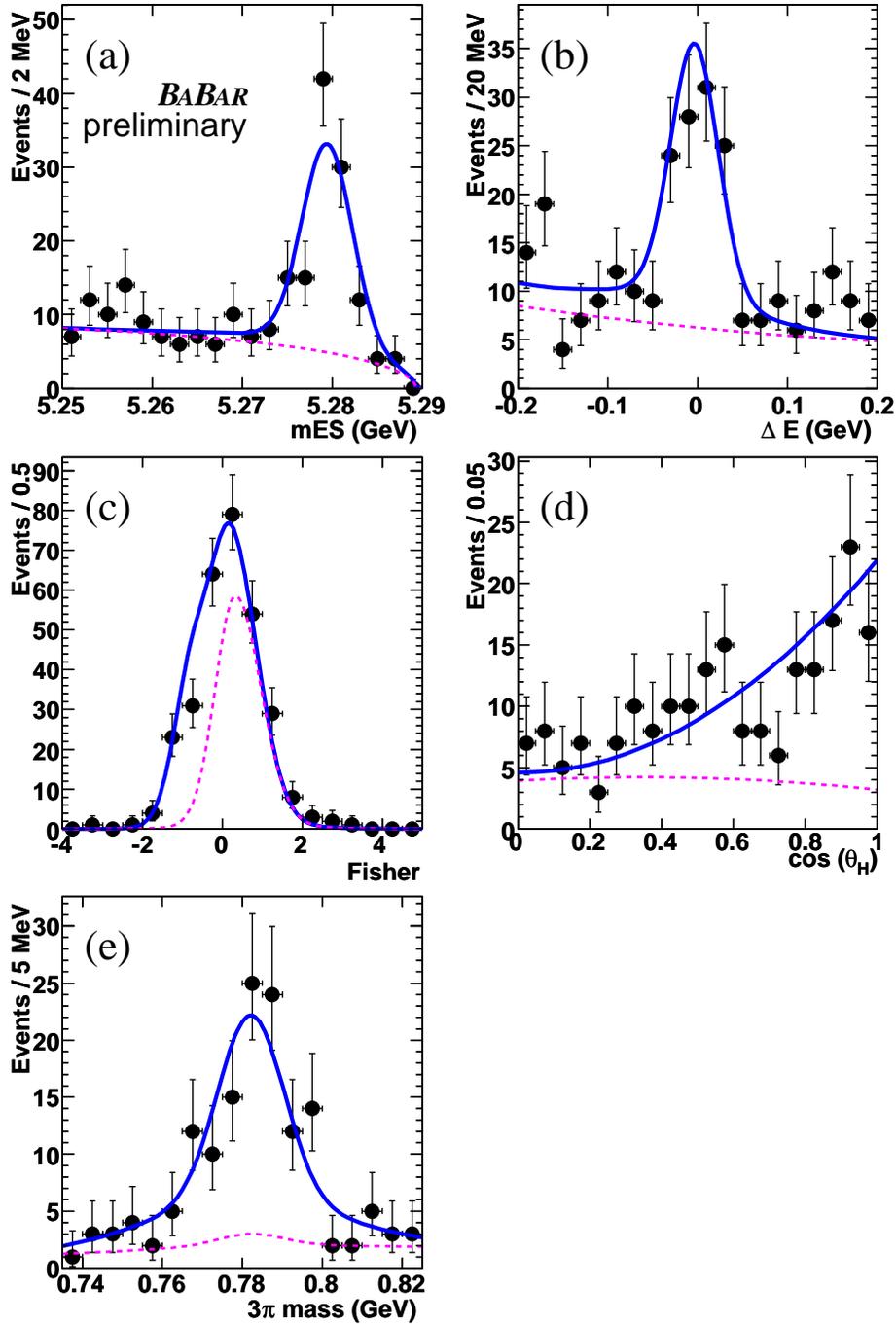}
\vspace{-.4cm}
 \caption{\label{fig:projMbDE}
$B$ candidate projections for \omegaKs\ of (a) \mb, (b) \DE,
(c) \xf, (d) \hel, and (e) \mres, shown for a 
signal-enhanced subset of the data (points with error bars), with
the fit function (solid line), and the background components (dashed line)
overlaid.}
\vspace{-.4cm}
\end{center}
\end{figure}

\begin{figure}[!tbp]
\begin{center}
\includegraphics[scale=0.7]{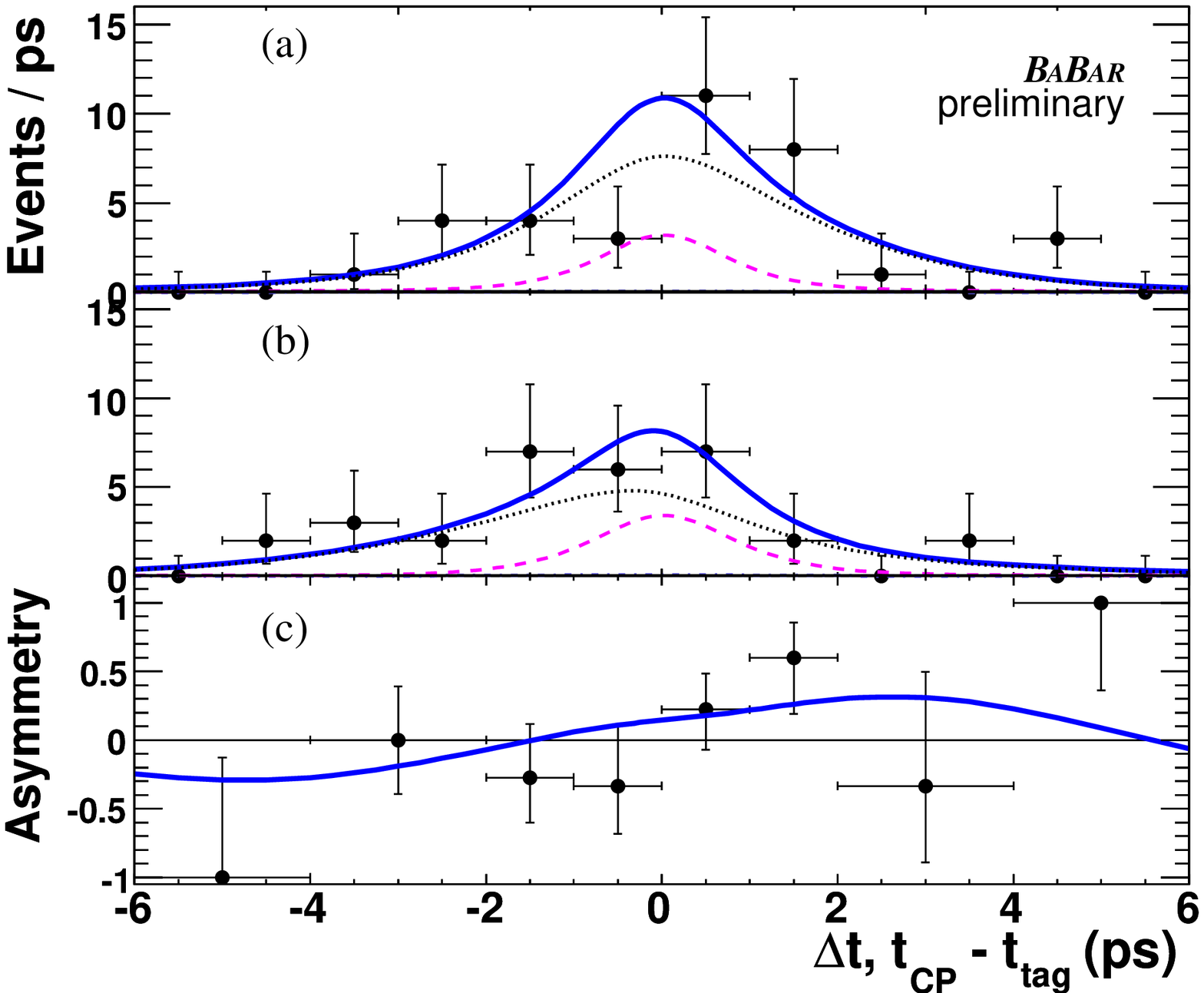}
\end{center}
\vspace{-.3cm}
\caption{Projections onto \deltat\ for \omegaKs, where $t_{CP}$ is the 
decay time for the signal $B$ meson.  Data (points with errors),
the fit function (solid line), background component (dashed line),
and signal component (dotted line) are shown for 
events in which the tag meson is
(a) \Bz\ and (b) \Bzb, and the
asymmetry $(N_{B^0}-N_{\Bbar^0})/(N_{B^0}+N_{\Bbar^0})$ is shown in (c), 
where $N$ indicates the total number of events passing the same cuts as for 
Fig.~\ref{fig:projMbDE}.}
\label{fig:dtproj}
\vspace{-0.4cm}
\end{figure}

Fig.\ \ref{fig:projMbDE}\ shows projections onto the fit variables for
a subset of the data (including 45--65\% of signal events) 
for which the signal likelihood
(computed without the variable plotted) exceeds a 
threshold that optimizes the sensitivity. Fig.~\ref{fig:dtproj} shows the 
$\Delta t$ projections and asymmetry of the time-dependent fit applying the
same event selection criteria as for Fig.~\ref{fig:projMbDE}. Based on
explicit variation of $C$ with $S$ allowed to float, we find the 
correlation between $S$ and $C$ to be negligible.

\section{SYSTEMATIC UNCERTAINTIES}

We estimate systematic uncertainties in $S$ and $C$ from the following
sources: potential dilution
due to \BB\ background (0.01); variation of the PDF shapes used in the
fit (0.01); knowledge of the parameters used to model the signal 
\dt\ distribution (0.02); and interference between the 
CKM-suppressed $\bar{b}\to\bar{u} c\bar{d}$ amplitude and the favored 
$b\to c\bar{u}d$ amplitude for some tag-side $B$ decays \cite{dcsd}
(0.02 for $C$, negligible for $S$), where the value in parentheses is the
size of the estimated systematic uncertainty.  
By applying distortions to MC samples 
and refitting all tracks, we find that the uncertainty due
to possible SVT misalignment and position and size of the beam spot 
are negligible.  The uncertainties in the parameters of fits to the \bflav\ 
sample are used for the uncertainties in the signal 
PDF parameters: \dt\ resolutions, tagging efficiencies, and mistag rates. 
Published measurements \cite{PDG2006} are used for $\tau_B$ and \deltamd.  
Summing all systematic uncertainties in quadrature, we obtain 0.02 for $S$ and
0.03 for $C$.

\section{SUMMARY}
\label{sec:Summary}

In conclusion, we have presented preliminary results for
the time-dependent asymmetry parameters for the decay \omegaKs,
$\skz = \SomegaKz$ and $\ckz = \ComegaKz$, where the first uncertainty is 
statistical and the second systematic.  If we fix
$C=0$, we find $S=0.63^{+0.28}_{-0.33}$, where the uncertainty is statistical
only.
This value of $\skz$ and the world-average value of
\stwob\ \cite{s2b,belles2b} yield a value of $\deltaS=-0.09\pm0.31$, in
good agreement with the Standard Model expectation near zero.

\section{ACKNOWLEDGMENTS}
\label{sec:Acknowledgments}

We are grateful for the 
extraordinary contributions of our \pep2\ colleagues in
achieving the excellent luminosity and machine conditions
that have made this work possible.
The success of this project also relies critically on the 
expertise and dedication of the computing organizations that 
support \babar.
The collaborating institutions wish to thank 
SLAC for its support and the kind hospitality extended to them. 
This work is supported by the
US Department of Energy
and National Science Foundation, the
Natural Sciences and Engineering Research Council (Canada),
Institute of High Energy Physics (China), the
Commissariat \`a l'Energie Atomique and
Institut National de Physique Nucl\'eaire et de Physique des Particules
(France), the
Bundesministerium f\"ur Bildung und Forschung and
Deutsche Forschungsgemeinschaft
(Germany), the
Istituto Nazionale di Fisica Nucleare (Italy),
the Foundation for Fundamental Research on Matter (The Netherlands),
the Research Council of Norway, the
Ministry of Science and Technology of the Russian Federation, 
Ministerio de Educaci\'on y Ciencia (Spain), and the
Particle Physics and Astronomy Research Council (United Kingdom). 
Individuals have received support from 
the Marie-Curie IEF program (European Union) and
the A. P. Sloan Foundation.


\begin{thebibliography}{99}

\bibitem{s2b}
\babar\ Collaboration, B.~Aubert \etal, \jprl{94}, 161803 (2005).
\bibitem{belles2b}
  Belle Collaboration, K.~Abe \etal, \jprd{66}, 071102(R) (2002).
\bibitem{SM}
N.~Cabibbo, \jprl{10}, 531 (1963); M.~Kobayashi and T.~Maskawa, \progtp{49}, 652
 (1973).
\bibitem{Penguin}
Y.~Grossman and M.~P.~Worah, \plb{395}, 241 (1997);
D.~Atwood and A.~Soni, \plb{405}, 150 (1997).

\bibitem{Gross} Y.~Grossman \etal, \jprd{68}, 015004 (2003).
\bibitem{BN} M.~Beneke and M.~Neubert, \npb{675}, 333 (2003).
\bibitem{beneke} M.~Beneke, \plb{620}, 143 (2005).
\bibitem{CCS} H-Y.~Cheng, C-K.~Chua, and A.~Soni, \jprd{72}, 014006 (2005).
\bibitem{lonsoni} D.~London and A.~Soni, \plb{407}, 61 (1997).

\bibitem{BABAR} \babar\ Collaboration, B.~Aubert \etal, \jprd{74}, 011106 (2006).

\bibitem{BELLE}
Belle Collaboration, K.~F.~Chen \etal, \jprd{72}, 012004 (2005).

\bibitem{BABARNIM}
\babar\ Collaboration, B.~Aubert \etal, \nima{479}, 1 (2002).

\bibitem{geant}
The \babar\ detector Monte Carlo simulation is based on GEANT4:
S.~Agostinelli \etal, \nima{506}, 250 (2003).

\bibitem{PRD}
\babar\ Collaboration, B.~Aubert \etal, \jprd{70}, 032006 (2004).

\bibitem{argus}
$f(x) \propto x\sqrt{1-x^2}\exp{\left[-\xi(1-x^2)\right]}$, with 
$x\equiv\mes/E_B^*$ and $\xi$ a parameter to be fit.
See ARGUS Collaboration, H.\ Albrecht \etal, \plb{241}, 278 (1990).

\bibitem{PDG2006}
Particle Data Group, Y.-M. Yao et al., \jpg{33}, 1 (2006).

\bibitem{dcsd}
O. Long, M. Baak, R. N. Cahn, and D. Kirkby, \jprd{68}, 034010
(2003).

\bibitem{BelleCLEO}
Belle Collaboration, C.H. Wang \etal, \jprd{70}, 012001 (2004);
CLEO Collaboration, C.P. Jessop \etal, \jprl{85}, 2881 (2000).

\bibitem{Previous}
\babar\ Collaboration, B.~Aubert \etal, \jprl{92}, 061801 (2004).

\end{thebibliography}
\end{document}